\pdfoutput=1
\makeatletter
\declare@file@substitution{revtex4-1.cls}{revtex4-2.cls}
\makeatother

\documentclass{aastex631}
\usepackage{graphicx} % Required for inserting images
\usepackage{amsmath}

\begin{document}

\title{Observation-Based Iterative Map for Solar Cycles. I. Nature of Solar Cycle Variability}
\correspondingauthor{Jie Jiang}
\email{jiejiang@buaa.edu.cn}

\author{Zi-Fan Wang}
\affiliation{State Key Laboratory of Solar Activity and Space Weather, National Astronomical Observatories, Chinese Academy of Sciences, Beijing 100101, China}
\affiliation{School of Astronomy and Space Science, University of Chinese Academy of Sciences, Beijing, China}

\author[0000-0001-5002-0577]{Jie Jiang}
\affiliation{School of Space and Earth Sciences, Beihang University, Beijing, China}
\affiliation{Key Laboratory of Space Environment Monitoring and Information Processing of MIIT, Beijing, China}

\author{Jing-Xiu Wang}

\affiliation{State Key Laboratory of Solar Activity and Space Weather, National Astronomical Observatories, Chinese Academy of Sciences, Beijing 100101, China}
\affiliation{School of Astronomy and Space Science, University of Chinese Academy of Sciences, Beijing, China}
\begin{abstract}
    %Inter-cycle variations in the series of 11-year solar activity cycles have a significant impact on both the space environment and climate. The dominance of either deterministic chaos or stochastic perturbations in solar cycle variability remains an open question. Distinguishing between the two mechanisms is crucial for the predictability of solar cycles. Here we reduce the solar dynamo process responsible for the solar cycle to a one-dimensional iterative map, incorporating recent advance in the observed nonlinearity and stochasticity of the cycle. We demonstrate that it naturally eliminates chaotic behavior when the generation of the poloidal field follows an increase-then-saturate pattern as the cycle strength increases. The result is supported by the stability of the fixed point and the cobweb diagrams, implying that stochasticity is the primary source of solar cycle variability.  Using the iterative map, we can generate a series of cycle amplitudes comparable to observations. The uncertainty of nonlinearity and stochasticity parameters affects the probability distribution of cycle amplitude, but cannot bring the system to chaotic states. Our results provide a guideline to the analysis of solar dynamo models in terms of chaos and stochasticity, offer an limitation to cycle prediction range, and also motivate further improvement of the observational constraint of nonlinear and stochastic processes.
    Inter-cycle variations in the series of 11-year solar activity cycles have a significant impact on both the space environment and climate. Whether solar cycle variability is dominated by deterministic chaos or stochastic perturbations remains an open question. Distinguishing between the two mechanisms is crucial for predicting solar cycles. Here we reduce the solar dynamo process responsible for the solar cycle to a one-dimensional iterative map, incorporating recent advance in the observed nonlinearity and stochasticity of the cycle. We demonstrate that deterministic chaos is absent in the nonlinear system, regardless of model parameters, if the generation of the poloidal field follows an increase-then-saturate pattern as the cycle strength increases. The synthesized solar cycles generated by the iterative map exhibit a probability density function (PDF) similar to that of observed normal cycles, supporting the dominant role of stochasticity in solar cycle variability. The parameters governing nonlinearity and stochasticity profoundly influence the PDF. The iterative map provides a quick and effective tool for predicting the range, including uncertainty, of the subsequent cycle strength when an ongoing cycle amplitude is known. Due to stochasticity, a solar cycle loses almost all its original information within 1 or 2 cycles. Although the simplicity of the iterative map, the behaviors it exhibits are generic for the nonlinear system. Our results provide guidelines for analyzing solar dynamo models in terms of chaos and stochasticity, highlight the limitation in predicting solar cycle, and motivate further refinement of observational constraints on nonlinear and stochastic processes.
\end{abstract}
\section{Introduction}\label{sec:intro}
%Solar cycle variability
The 11-year solar activity cycle is the most prominent feature of the solar magnetic field \citep{Hathaway2015}.  Usually indicated by the peak sunspot number during one cycle, the cycle amplitudes vary within a vast range, including periods of very low activity such as the Maunder Minimum \citep{1976Sci...192.1189E}, and periods of high activity such as the Modern Maximum \citep{1995GeoRL..22.3195L}.  Moreover, the normal cycles exhibit a weak-strong alternating tendency, known as the Gnevyshev-Ohl rule (G-O rule, Even-Odd effect) \citep{1948G_O}.  These variations shape the observed solar cycle, which may seem random but reveals key semi-regular patterns. These patterns offer valuable insights into cycle prediction and space climate research.

It is widely accepted that a solar dynamo process is responsible for the origin and evolution of solar cycles \citep{2020LRSP...17....4C}.  The nonlinearity and stochasticity of the solar dynamo are key to explaining cycle variability, including the range of cycle amplitudes and semi-regular patterns like the G-O rule.  Solar dynamo explains the solar cycle as the mutual generation of toroidal magnetic field and poloidal field \citep{1955ApJ...122..293P,1966ZNatA..21..369S,1979cmft.book.....P}.  The solar dynamo is evident to be of Babcock-Leighton (B-L) type \citep{Cameron2015},  in which the generation of toroidal field from poloidal field is generally linear.  The linear relationship is supported by the polar field precursor for cycle prediction as the polar field at cycle minimum is taken as the representative of the poloidal field \citep{1979stp.....2..258O,2005GeoRL..32.1104S,2005GeoRL..3221106S}.  Besides the linear part, solar dynamo also contains nonlinear mechanisms to confine the amplitude range \citep{1978ApJ...220..692Y}, making the next cycle weaker for strong cycles, similar to the G-O rule.  Earlier studies suggested that certain nonlinear feedback mechanisms can cause dynamo models to undergo a clear transition to chaos, marked by a sequence of period-doubling bifurcations \citep{1978ApJ...221.1088Y,1991A&A...245..654S,2005ApJ...619..613C}.  Such deterministic chaos may explain the irregularities of solar cycle variations.  Meanwhile, as the generation of poloidal field from toroidal field happens in the convection zone \citep{1955ApJ...122..293P}, the stochasticity arising from turbulent convection also influences the solar cycle evolution, probably generating irregular cycle amplitude variations.

Whether deterministic chaos or stochasticity causes the solar cycle variability is yet to be answered. Some attempts on analysis of time series of solar activity proxies \citep{Mundt1991,Rozelot1995,Hanslmeier2010,Deng2016} or investigations based on dynamic equations  \citep{Tobias1995,Knobloch1998, Charbonneau2001,Wilmot-Smith2005} suggest that solar cycle variability is due to deterministic chaos subject to weak stochastic perturbations. In contrast, with similar methods, some studies \citep{Price1992, Carbonell1994, Mininni2000, Kitchatinov2016, Cameron2017} demonstrated that solar cycle variability results from weakly nonlinear limit cycle affected by stochastic perturbations. Understanding the nature of solar cycle variability is a prerequisite for reasonably predicting future solar cycles \citep{Petrovay2020}, and is also the prototype for understanding stellar variability.
%Solar dynamo and its nonlinearity, stochasticity

While certain properties of solar dynamo and the nature of cycle variability can be evaluated by kinematic or even fully magnetohydrodynamic simulations \citep{2020LRSP...17....4C, Karak2023}, simplifying the dynamo into an iterative map of cycle amplitudes provides an effective and convenient method for evaluating their properties in the presence of dominant chaos or stochasticity, as long as the nonlinearity and stochasticity are realistically quantified \citep{May1976, 2005ApJ...619..613C}.
Pioneered by \citet{Durney2000} and \citet{Charbonneau2001}, iterative maps of solar cycles are not as developed as other methods, but more recent progress on solar dynamo can revive such iteration map studies.

Within the B-L type dynamo, the generation of poloidal field from toroidal field results from the emergence of tilted active regions and their subsequent surface flux transport evolutions \citep{Wang1991,2013A&A...553A.128J,Petrovay2020JSWSC,2023SSRv..219...31Y}.  Such surface evolution process can be observationally confined.  Studies on historical datasets indicate that with increasing cycle strength, the average tilt angle decreases \citep{2010A&A...518A...7D, 2021A&A...653A..27J} and the average latitude increases \citep{2003SoPh..215...99L, 2008A&A...483..623S}, which consequently limits the generation of poloidal field. This B-L type nonlinearity can effectively modulate the polar field generation \citep{Jiang2020, Karak2020, Talafha2022, Yeates2024}. Meanwhile, as the rise of toroidal flux through the convection zone to form active regions is buffeted by the turbulent convection \citep{Weber2013}, the latitude and tilt are always of large scatter \citep{2011A&A...528A..82J, Jiang2014}. This observation-based B-L stochasticity can modulate solar cycles, the relative importance of which has been evaluated recently \citep{Jiang2020, Talafha2022,2024MNRAS.530.2895K}.

The observations and theoretical studies of B-L dynamo make it possible to create a realistic recursion relation between the amplitude of adjacent cycles.  Following the quantification of B-L type nonlinearity and stochasticity by \citet{Jiang2020} (hereafter J20), we create an iterative map of solar cycle amplitudes.  In this paper, as the first of the iterative map series, we perform an analysis of the origin of solar cycle variability, answer whether deterministic chaos or stochasticity dominates the variability, and investigate how nonlinearity and stochasticity of different strength affect the variability, by evaluating the properties of the solar cycle iterative map.  We will continue evaluating the Gnevyshev-Ohl rule using the iterative map in a subsequent paper. In the third paper, we will first validate the iterative map for solar cycle prediction and then apply it to predict the range of solar cycle 26, based on the amplitude of cycle 25.

%Here we present an analysis of the origin of solar cycle variability, answer whether deterministic chaos or stochasticity dominates the variability, and investigate how nonlinearity and stochasticity of different strength affect the variability, by creating and analyzing an iteration map of solar cycles based on B-L dynamo.  B-L dynamo itself implies a recursion relation between two adjacent cycles via the evolution of polar fields.  The B-L nonlinearity and stochasticity can be well parameterized following \citet{Jiang2020}'s (hereafter J20) results.  Following these, we create the iteration map and then show the role of nonlinearity and stochasticity in cycle variability.

The paper is organized as follows.  In Section \ref{sec:method} we construct the iterative map and describe its parameters.  In Section \ref{sec:result} we analyze the origin of cycle variability and the distribution of cycle amplitudes from series of cycle amplitudes synthesised from the iterative map.  We discuss the implication for cycle prediction of our results in Section \ref{sec:disc}.  We conclude our paper in Section \ref{sec:outro}.

%Observational nonlinearity and stochasticity in B-L dynamo

%Iteration map of cycle amplitude to determine the role of dynamo ingredients and examine cycle variation properties

\section{Iterative map of solar cycle amplitudes}\label{sec:method}

The B-L type dynamo offers a framework for establishing a relationship between successive solar cycles, in which the generation of the poloidal field from toroidal field can be constrained by observations.  \citet{Durney2000} and \citet{Charbonneau2001} first constructed a nonlinear recursion function of solar cycle amplitude based on the B-L type dynamo.  We adopt this methodology, incorporating recent understandings of the B-L mechanism, particularly those presented by J20, to create the 1D iterative map for cycle-to-cycle amplitude variations. The map enables us to explore the nature the solar cycle variability using established tools from nonlinear dynamics, such as fix-point stability and cobweb diagrams.

The solar dynamo process is composed of the mutual generation of poloidal and toroidal field.  The poloidal field at cycle minimum is crucial in B-L dynamo for being the seed of the following cycle.  We use its analog, the axial dipole moment of surface field at cycle minimum, which is denoted as $DM\left(n\right)$ for the dipole moment at the end of $n$-th cycle.  As for the toroidal field during the solar cycle, it can be represented as the cycle amplitude which describes the active region emergence during the cycle, denoted as $SN\left(n\right)$ for the $n$-th cycle.

We begin with the first assumption that the generation of toroidal field, i.e. the $\Omega$-effect, is mostly linear and less noisy compared to other processes.  This corresponds to the polar field precursor mentioned in Section \ref{sec:intro}.  This is written as follows,

\begin{equation}\label{eq:omega}
    SN\left(n\right) = k_{0}DM\left(n-1\right),
\end{equation}
in which $k_{0}$ is the linear coefficient that describes the strength of the $\Omega$-effect and can be derived from the observed solar cycle amplitude and the surface magnetic field at cycle minima.

We then consider how the poloidal field is in turn generated from the toroidal field.  Originally, \citet{Durney2000} suggested that the newly generated poloidal field, or $DM\left(n\right)$ in our notation, is a certain function of the toroidal field, or $SN\left(n\right)$ in our notation.  However, as the B-L mechanism suggests, during the solar cycle $n$, originating from active regions, the surplus flux of opposite sign to the polar field migrates poleward, causes polar field reversal before building up the polar field at the next cycle minimum.  This imply that, instead of directly generating $DM\left(n\right)$, the total toroidal field during cycle $n$, represented by $SN\left(n\right)$, is actually generating the difference between adjacent $DM$s, as shown by the following,

\begin{equation}
    \Delta DM (n) = DM\left(n\right) + DM\left(n-1\right) ,
    \label{eq:ddm}
\end{equation}
where $\Delta DM (n)$ is a function of $SN\left(n\right)$, representing the total amount of dipole moment at cycle minimum contributed by all the active regions during cycle $n$.  The values of $DM$ and $SN$ we use in this article are all positive, and thus it is the positive sign in the equation.  Here we include the second assumption that Hale's polarity law is always kept, so that the new dipole moment always cancels out with the old.

Combining Equations (\ref{eq:omega}) and (\ref{eq:ddm}), a complete recursion relation between adjacent cycles can be obtained if a certain form of $\Delta DM$ is given.  Unlike \citet{Durney2000} and \citet{Charbonneau2001}, who introduced a parameter to represent the strength of the nonlinearity based on implication of rising thin flux tube modeling, we base our approach on the results of J20, which offer an observation-based, clearer interpretation and quantification of the nonlinearity in the B-L dynamo.  Using empirical rules of AR emergence, and surface flux transport simulations, J20 shows that $\Delta DM$ follows a nonlinear function of $SN\left(n\right)$. J20 includes tilt quenching and latitude quenching as B-L nonlinearity, and the scatter of active region properties as B-L stochasticity. $\Delta DM$ follows a increase-then-saturate pattern as the increase of cycle amplitude, which means that the generation of poloidal field increases when cycle amplitude increases from weak to strong, but does not increase or decrease when the cycle becomes even stronger. This is different from increase-then-decrease pattern, such as the results of \citet{Charbonneau2001} where there is a decreasing part after the cycle becomes too strong. The saturated pattern of J20 primarily arises from the similar decay properties of solar cycles in sunspot numbers \citep{Cameron2016, 2018ApJ...863..159J, Biswas2022}.

The J20 form of $\Delta DM$ is written below, with a deterministic case including only nonlinearity and a stochastic case including both nonlinearity and stochasticity.
\begin{equation}\label{eq:nlin}
    \Delta DM =\begin{cases}
         k_{1}\textbf{erf}\left(\frac{SN\left(n\right)}{quench}\right) & \text{deterministic case;}\\
         k_{1}\textbf{erf}\left(\frac{SN\left(n\right)}{quench}\right)\left(1 + stoch\times X\right) & \text{stochastic case,}
    \end{cases}
\end{equation}
where $\textbf{erf}$ is the error function. The parameter $k_1$ represents the saturated value of the total dipole moment $\Delta DM$ contributed by all ARs during a strong solar cycle. The parameter $quench$ determines the rate at which $\Delta DM$ saturates as a function of sunspot number $SN\left(n\right)$ of cycle $n$. Finally, the parameter $X$ is a normally distributed random variable with a standard deviation given by $stoch$. Figure 4 of J20 illustrates the dependence of $\Delta DM$ on $SN\left(n\right)$.

With the realistic $\Delta DM$, the recursion function of cycle amplitudes can be finalized.  The deterministic case is as follows,
\begin{equation}
    SN\left(n+1\right) = k_{0}k_{1}\textbf{erf}\left(\frac{SN\left(n\right)}{quench}\right) - SN\left(n\right),
    \label{eq:recursion}
\end{equation}
while the stochastic case is
\begin{equation}
    SN\left(n+1\right) = k_{0}k_{1}\textbf{erf}\left(\frac{SN\left(n\right)}{quench}\right)\left(1 + stoch\times X\right) - SN\left(n\right).
    \label{eq:recursion2}
\end{equation}
These two cases can be used to generate series of solar cycle amplitudes for analysis efficiently when an initial condition is given.  Meanwhile, they can also be used to predict solar cycles with a given $SN\left(n\right)$, including uncertainties of prediction.

The parameters in Equation (\ref{eq:recursion2}) are obtained from observations and surface flux transport simulations. The parameter $k_{0}= 58.7$ \citep{2018ApJ...863..159J} for the linear $\Omega$-effect is derived from the linear relationship between the axial dipole moment (based on WSO, MWO, MDI/SOHO, and HMI/SDO synoptic magnetograms at the cycle $n-1/n$ minimum) and the maximum value of the 13-month smoothed monthly sunspot number over cycle $n$ in Sunspot Number Version 2. For the nonlinear and stochastic B-L mechanism, the parameters need to be obtained by simulating the surface flux transport of ARs following observational properties, like J20 did.  We adopt the parameters of J20, with maximum dipole moment $k_{1}$ being 6.94, $quench$ being 75.85, and $stoch$ being 0.17, in order to analyze the nature of cycle variability.  This parameter set is referred to as the standard set.

We note that the cycle amplitude $SN(n)$ used in this paper is the widely adopted maximum value of the 13-month smoothed monthly sunspot number for cycle $n$ \citep{Hathaway2015}, rather than an integrated value. The maximum value of sunspot number is approximately proportional to the total sunspot area \citep{Balmaceda2009} and total magnetic flux \citep{Jiang2011b}, which are directly related to the toroidal field during a solar cycle. While the detailed evolution of the cycle may slightly influence the total toroidal field, we do not account for these variations in our current approximation.

The parameters have uncertainty because of the limitations of observational and statistical study of ARs and surface flows.  The uncertainty of ARs and flows then affect the production of $\Delta DM$ by surface flux transport processes \citep{2014SSRv..186..491J, 2023SSRv..219...31Y}, and should be examined with numerical simulations.  In this paper, we do not examine the quantity of model parameter uncertainty caused by uncertainty of ARs and flows.  Instead, we introduce a considerable range of parameters and study their influence on the solar cycle variability.

%The parameters have uncertainty due to the limitations of statistical analysis of ARs and surface flows.  As stated by Jiao et al. (2021) \cite{2021A&A...653A..27J}, the insufficient long-term data and difference in analysis methods have been causing mixed results of tilt quenching.  How does the uncertainty of tilt and latitude quenching as well as of the surface flows affect the model parameters should be examined with a surface flux transport model.

%In our work, we mainly focus on how different model parameters affect the variability and statistical properties of solar cycles.
%In our work, we introduce a certain range of the model parameters, and examine how the variability and statistical properties change.  We consider a $\pm 25\%$ variation of each parameter, changing one parameter at a time while keeping others unchanged.  6 cases are considered: $0.75\times k_{1}$, $1.25\times k_{1}$, $0.75\times quench$, $1.25\times quench$, $0.75\times stoch$, and $1.25\times stoch$.  The influence of parameter difference can then be obtained by comparisons of different series generated from the 6 parameter sets.

%If we have a relationship between the toroidal field and the newly generated poloidal field, we can produce a recursion relation of cycles, assuming that the Omega effect is generally linear.

%In this paper we use Jiang2020 results for the generation of new poloidal field.

%Considering that the parameters are subject to changes as we try to further our knowledge by more observations, we need to examine a wide range of parameters.

\section{Nature of solar cycle variability}\label{sec:result}

\subsection{No deterministic chaos in cycle variability}\label{subsec:variability}

With the recursion relation of $SN\left(n\right)$ originated from B-L dynamo, the role of nonlinearity and stochasticity in the solar cycle evolution and whether chaos is present can be evaluated by analyzing the recursion function and the synthesised solar cycle amplitude series.

\begin{figure}
%%\plotone{}
\gridline{\fig{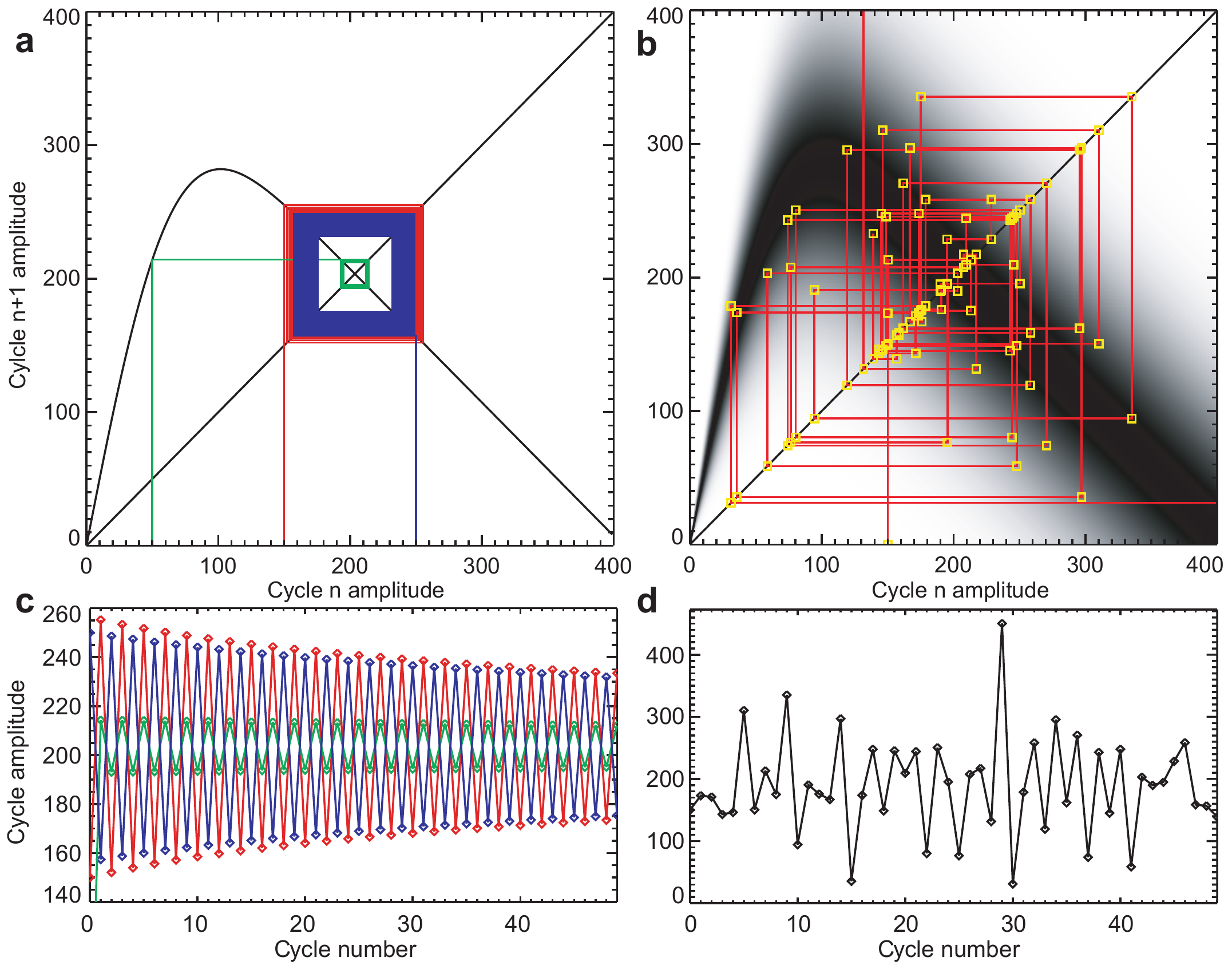}{0.9\textwidth}{}          }
\caption{Generation of solar cycle amplitudes with Cobweb diagrams.  {\bf a,} The Cobweb diagram for the deterministic case, with the black curve representing the recursion function Equation (\ref{eq:recursion}). The green, red, and blue curves represent the evolution of cycle amplitude with initial amplitudes of 50, 150, and 250, respectively. The corresponding cycle series is presented in Panel c.  The diagonal black line represents $SN\left(n\right)=SN\left(n+1\right)$.  {\bf b,} The Cobweb diagram for the stochastic case. The gray-scale shading represents the probability distribution of the iterative map, with deeper being more likely. Yellow squares indicate possible amplitudes of the cycle series, which are presented in Panel d.  {\bf c,} Cycle amplitude evolution for the deterministic case. The green, red, and blue curves are the evolution of cycle amplitude from Panel a.  {\bf d,} Cycle amplitude evolution for the stochastic case in Panel b.  \label{fig:cycs}}
\end{figure}

To determine whether determinisitic chaos is present, we first analyze the determinisic case without stochasticity.  The characteristics of the fixed point of the recursion function is critical to chaos.  We begin with analyzing the recursion function with the standard parameters, shown by the black curves in Figure \ref{fig:cycs} a.  As the panels show, the recursion function has 1 fixed points except the origin.  It is commonly known that the stability of a fixed point is determined by the derivative.  That is, let $f\left(x\right)$ be the recursion function of $\{x_{n}\}$, then a fixed point $x^{*}$ is stable if $\left|f^{'}\left(x^{*}\right)\right|<1$.  From Figure \ref{fig:cycs} a, the fixed point is on the decreasing part of the recursion function, so the derivative is obviously smaller than 1. Meanwhile, since the derivative of the deterministic case can be written as,

\begin{equation}\label{eq:dfdet}
    \frac{d SN\left(n+1\right)}{d SN\left(n\right)} = \frac{2k_{0}k_{1}}{\sqrt{\pi}quench}\exp\left(-\left(\frac{SN\left(n\right)}{quench}\right)^{2}\right) - 1,
\end{equation}
it can be clearly seen that it is always larger than -1.  Then we know that the fixed point has the absolute value of derivative smaller than 1, indicating that it is stable.  A stable fixed point results in converging into the fixed point instead of chaos for the deterministic case.  More generally, Equation (\ref{eq:dfdet}) follows,
\begin{equation}\label{ieq}
    \left|\frac{d SN\left(n+1\right)}{d SN\left(n\right)} \right|_{SN\left(n\right) = x^{*}} < 1,~~\forall~k_{0}, k_{1}, quench,
\end{equation}
which implies that the nonzero fixed point is stable regardless of parameters. Again, it is easy to know from Equation (\ref{eq:dfdet}) that $\frac{d SN\left(n+1\right)}{d SN\left(n\right)}>-1$.  Meanwhile, in order to have 1 nonzero fixed point, $\frac{d SN\left(n+1\right)}{d SN\left(n\right)}>1$ should hold at the origin.  Then, it can be easily know that $\frac{d SN\left(n+1\right)}{d SN\left(n\right)}<1$ at the nonzero fixed point, thus confirming Equation (\ref{ieq}).

From a larger perspective, this result is not only limited to $\Delta DM$ following Equation (\ref{eq:nlin}).  If $\Delta DM$ follows a increase-then-saturate form, which means that its derivative is large at the start, and gradually decreases, finally approaching the limit 0.  That is, $\frac{d \Delta DM}{d SN}>0$ while  $\frac{d^{2} \Delta DM}{d SN^{2}}<0$ if these derivatives exist.  For such kind of $\Delta DM$ with gradually decreasing derivative, the recursion function of cycles is concave and is easy to know that the nonzero fixed point has derivative smaller than 1 if the origin has derivative larger than 1 (we can assume that the derivative at the origin should be larger than 1 so that weak solar cycles would not converge to 0).  Meanwhile, the derivative of recursion function becomes closer to but always larger than -1 as the cycle amplitude increases.  Therefore, such a general kind of poloidal field generation mechanism has stable fixed point, which imply that they do not have chaos.

\begin{figure}
%%\plotone{}
\gridline{\fig{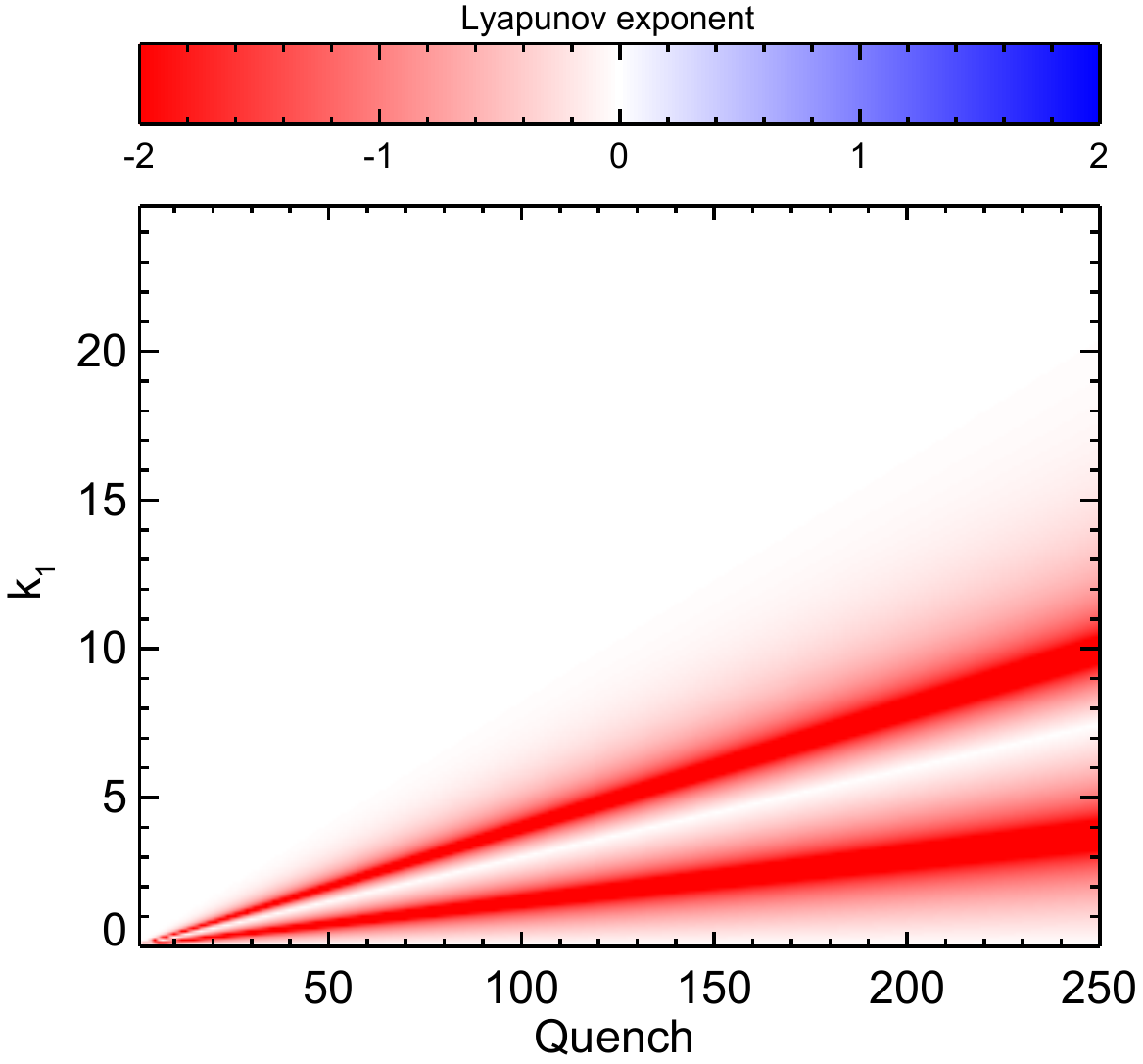}{0.5\textwidth}{}          }
\caption{Lyapunov exponents for various ranges of the parameters $quench$ and $k_1$.  \label{fig:lyap}}
\end{figure}

The absence of chaos implies that the initial uncertainty will not be enlarged exponentially.  This is characterized by the Lyapunov exponent, so we calculate the Lyapunov exponent of different $k_{1}$ and $quench$ for the iterative map following Equation (\ref{eq:recursion}).  The exponent is obtained from a series of 5000 cycles for each parameter set, with the initial cycle amplitude being 150.  The result is displayed by Figure \ref{fig:lyap}, which clearly presents 2 strips of negative values.  For larger $k_{1}$ and smaller $quench$ values, the exponent falls to almost 0, indicating extremely slow convergence to the fixed point.  Positive values are not observed in the figure, hence no chaos is present in the model for the parameter range we concern, that is, tiny difference between initial conditions will not grow unbounded.

To show the absence of chaos more clearly and intuitively, we generate series of cycle amplitudes and plot cobweb diagrams to show their iterations.   We generate series of the deterministic case as expressed by Equation (\ref{eq:recursion}).  The standard parameter set is used.  We consider 3 initial cycle amplitudes $SN\left(0\right) = 50$, 150, and 250, covering both below and above the fixed point where $SN\left(n+1\right) = SN\left(n\right)$ shown in Figure \ref{fig:cycs}a.  Then, the iterations starting from 3 initial conditions are shown with folding lines in cobweb diagrams along with the recursion functions and diagonal lines, as shown in Figures \ref{fig:cycs}a, respectively.

%The properties of the iteration function (Figure \ref{fig:cycs} a, b) and its properties -- its derivatives and stability of fixed point.

%The iteration shown in the cobweb shows no chaos, and stochasticity is required. (Figure \ref{fig:cycs} a-d)

%The Lyapunov exponent diagram shows that the deterministic case cannot have chaos regardless of parameters (Figure \ref{fig:lyap})

The 3 series of cycle amplitudes of the deterministic case are not chaotic. The corresponding cycle amplitude evolutions for the first 50 cycles are plotted in Figures \ref{fig:cycs}a, c.  All show regular alternating strong and weak patterns with amplitude slowly converging to the fixed point.  As shown in Figure \ref{fig:cycs}c, the convergence is slower for points closer to the fixed point.  The slow convergence results from Equation (\ref{eq:dfdet}) being only slightly larger than -1 near the fixed point, and also corresponds to the near 0 value of Lyanpuov exponent shown in Figure \ref{fig:lyap}.  The diagrams clearly illustrate that the deterministic case does not possess chaos.

Since chaos is not present in the model, stochasticity should be the primary source of cycle variability,  To show this, we generate a series of cycles and corresponding cobweb diagrams for the stochastic case.  We use recursion function Equation (\ref{eq:recursion2}) and initial amplitude 150, generate a series of cycle amplitudes, and plot the cycle evolution in Figures \ref{fig:cycs}b and \ref{fig:cycs}d.  The recursion function now exhibits a large scatter range compared to the deterministic case, and the iterations of the Cobweb diagram expanse across the displayed range.  The irregularities of cycle evolution are clear for the stochastic case, while converging behavior is not observed, different from the deterministic case.  The varying solar cycle amplitudes look similar to observations, suggesting that stochastic processes could be the main cause of solar cycle variability.

For the stochastic case, the predictability of the cycle amplitude is strongly limited by $k_{1}stoch$.  From Equation (\ref{eq:nlin}), the scatter of the next cycle for cycle $n$ is $k_{0}k_{1}stoch\textbf{erf}\left(\frac{SN\left(n\right)}{quench}\right)$, which is already notably diverging for the same $SN\left(n\right)$ as shown by the gray scale shade in Figure \ref{fig:cycs} (b).  Therefore, the predictability is always limited, and Lyapunov exponent is not needed for this case.  We will further discuss the predictability of solar cycles when stochasitcity is considered in Section \ref{sec:disc}.

\subsection{The probability density distribution of cycle amplitudes}\label{subsec:cycpdf}
%The "grand maxima" in usoskin2014 is about 257.75

%We want to obtain the PDF and compare with those reconstructed from cosmogenic isotopes.  First we need an conversion relation between Usoskin et al. 2014 and us. (Fig \ref{fig:uso})

It is important to examine whether the varying cycle amplitudes generated by the stochastic case follow realistic statistical properties.  We analyze the probability density distribution of the solar cycle amplitude produced by the stochastic case.  Using the standard parameter set, we generate 1,000,000 cycle amplitudes, and obtain their probability density function (PDF).  We then compare the results with the PDF of the long-term solar activity reconstructed from radioisotopes by \citet{2014A&A...562L..10U}.  However, before comparison, we need to convert \citet{2014A&A...562L..10U}'s results to match our convention.

\begin{figure}
%%\plotone{}
\gridline{\fig{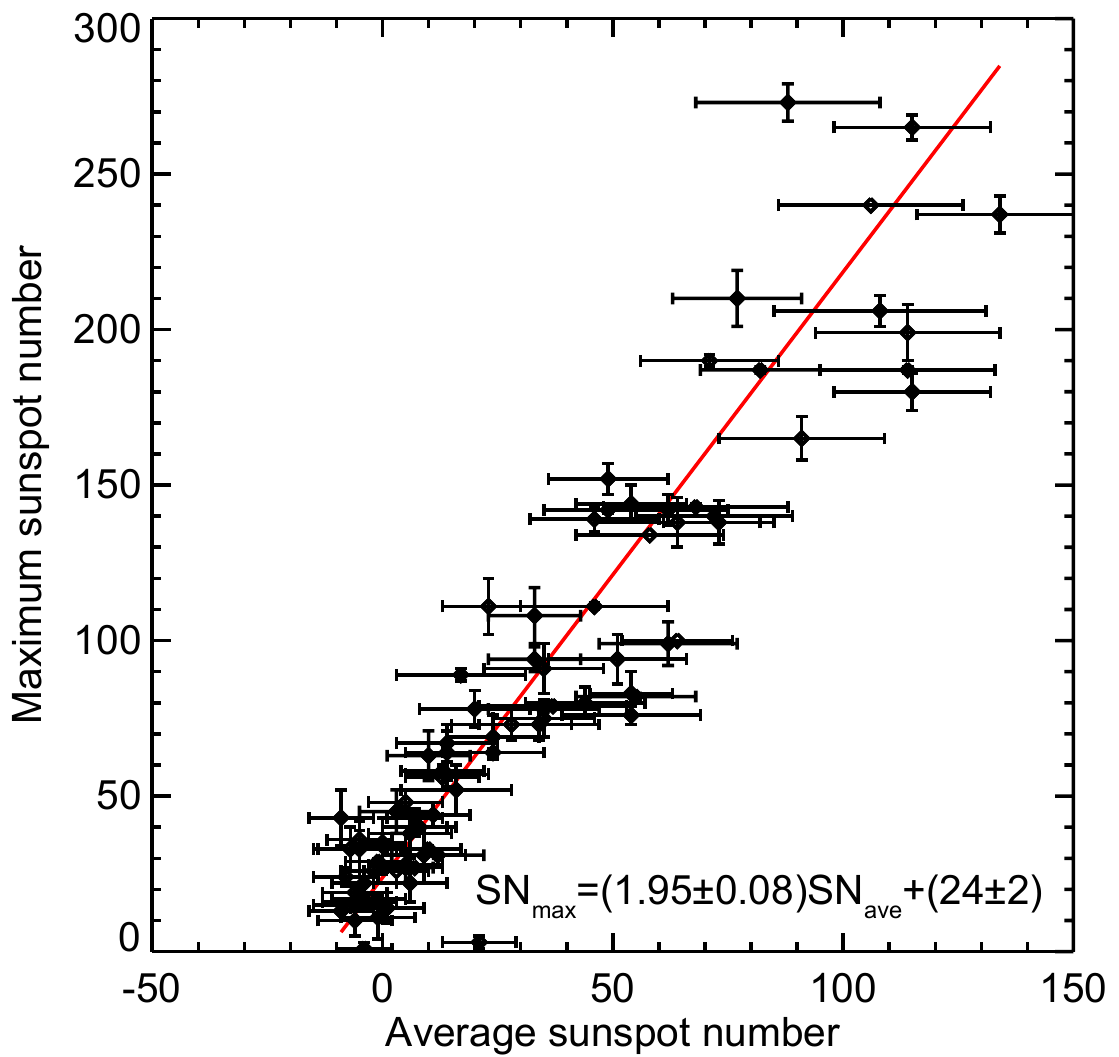}{0.5\textwidth}{}          }
\caption{Relationship between cycle average sunspot numbers and cycle maximum sunspot numbers. Each diamond represents a solar cycle, the error bars represent standard error, and the line represents linear fit. In total, 85 cycles occurred between the years 791 and 1900.  \label{fig:uso}}
\end{figure}

\citet{2014A&A...562L..10U} express the long-term solar activity variations as 10-year average of sunspot numbers, instead of the maximum sunspot number during a cycle.  Moreover, \citet{2014A&A...562L..10U} use a sunspot number series different from J20 which we mainly adopt.  To compensate these, we first need to find the relationship between cycle amplitude defined by the maximum sunspot number during a cycle and the 10-year average sunspot number.  To simplify the method, we assume that the 10-year average is analogous to solar cycle average.  \citet{2021A&A...649A.141U} contains resolved solar cycles in the recent millennium.  We find the cycle maximum sunspot number (the cycle amplitude we use) and the cycle average sunspot number from their results, as well as the corresponding measure errors.   There are 85 cycles in total, shown in Figure \ref{fig:uso}.  The two sunspot number series follow a linear relationship, so we perform a linear fit \citep{2009ASPC..411..251M}.  The relationship between average sunspot number $SN_{ave}$ and maximum sunspot number $SN_{max}$ is then $SN_{max}=\left(1.95\pm0.08\right)SN_{ave}+\left(24\pm2\right)$.  The $\chi^{2}$ of the fit is 94.

The cycle amplitudes of \citet{2014A&A...562L..10U} is compared to the Group Sunspot Number series \citep{1998SoPh..179..189H}, which is scaled to the Wolf Sunspot Number since 1874.  The version 2.0 of the International Sunspot Number can be regarded as the Wolf Sunspot Number divided by 0.6 \citep{2014SSRv..186...35C}.  Hence, we also use a factor of $1/0.6$ when we compare our results with the results of \citet{2014A&A...562L..10U}.  The relationship is as follows,
\begin{equation}
SN=\left[\left(1.95\pm0.08\right)SN_{Usoskin}+\left(24\pm2\right)\right]/0.6,
\end{equation}
and we use this to convert the data when we compare the PDF of the model with observations.

\begin{figure}
%%\plotone{}
\gridline{\fig{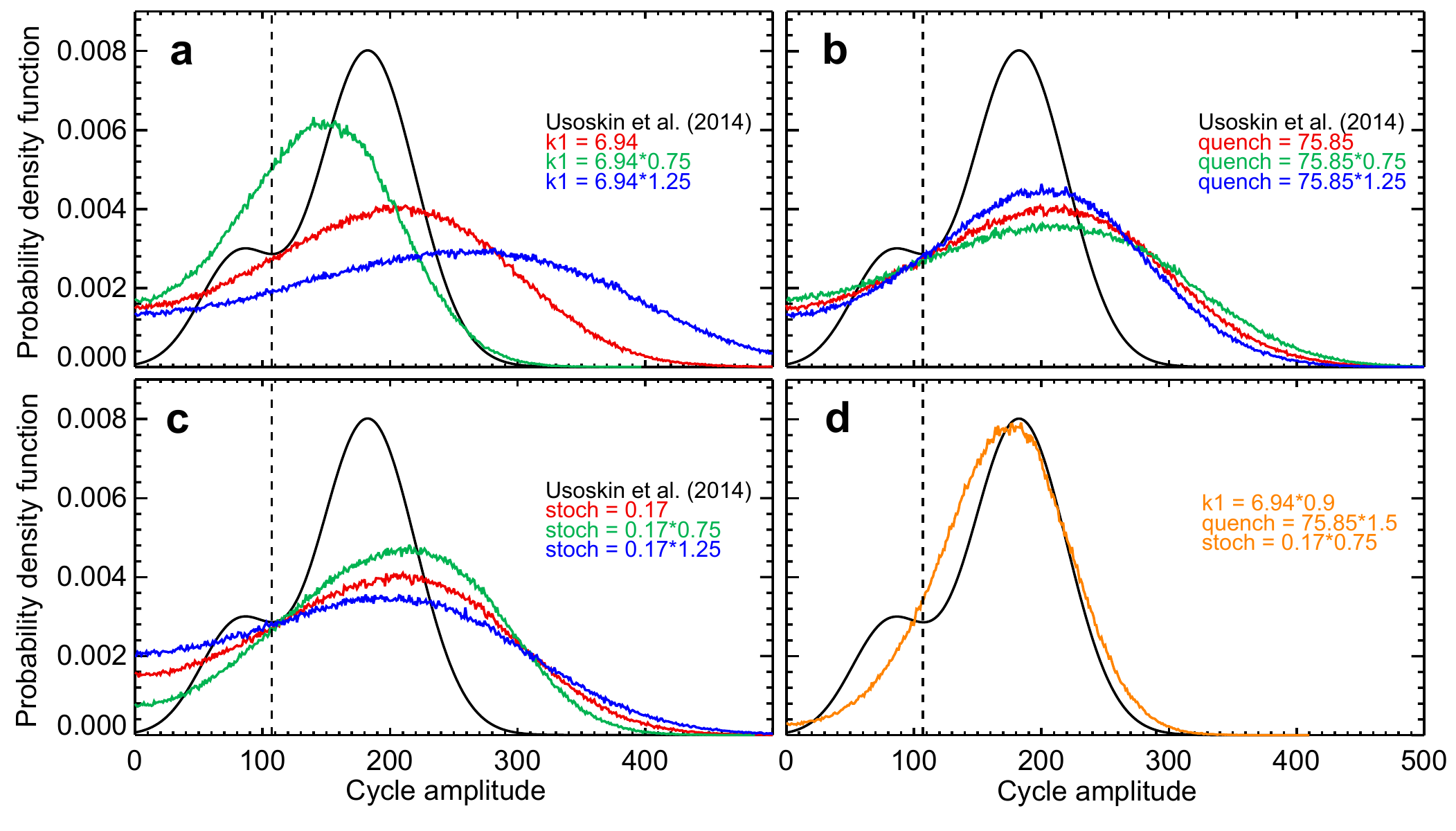}{1\textwidth}{}          }
\caption{PDFs of cycle amplitude, under different parameter sets.  {\bf a-c,} Modified $k_{1}$, $quench$, and $stoch$ respectively, with different colors displaying different parameters as stated in the panels.  {\bf d,} A case that modifies both $k_{1}$ and $quench$, displayed in orange.  The black curves are from \citet{2014A&A...562L..10U}.  The black dashed vertical line marks the weakest cycle amplitude 107.1 to deduce the empirical rules of ARs in J20.  \label{fig:pdfs}}
\end{figure}

Using the stochastic case of recursion function, we generate 1,000,000 solar cycle amplitudes and obtain the corresponding PDF, shown by the red curve in Figures \ref{fig:pdfs} a-c.  Since our method is based on the empirical properties of the ARs used in J20, only the amplitudes larger than the weakest observed cycle used in J20 (cycle 14 with amplitude 107.1) are constrained by observations, marked by the part to the right of the dashed line.  Qualitatively, the PDF curves consist of a single peak similar to the normal cycle component in the PDF of \citet{2014A&A...562L..10U} shown in black.  However, at low cycle amplitudes to the left of the dashed line, i.e., grand minima, the curves differ from \citet{2014A&A...562L..10U} because they do not have the grand minima component.  Quantitatively, the peak amplitude is approximately 200, which is within 11\% of the reconstructed normal cycle component in the PDF of \citet{2014A&A...562L..10U}.  The peak probability density and distribution width are less consistent with the reconstructed results, though.  The quantitative difference to observations is likely due to the uncertainty of nonlinearity and stochasticity parameters.

The PDF of synthetic cycle amplitude varies with nonlinearity and stochasticity parameters.  We consider a $\pm 25\%$ variation of each parameter, changing one parameter at a time while keeping others unchanged.  Thus 6 cases are considered: $0.75\times k_{1}$, $1.25\times k_{1}$, $0.75\times quench$, $1.25\times quench$, $0.75\times stoch$, and $1.25\times stoch$.  The influence of parameter difference can then be obtained by comparisons of different series generated from the 6 parameter sets.

The PDFs of the 6 parameter sets are shown in different colors in Figures \ref{fig:pdfs}a-c.  Among the 3 parameters, the maximum dipole moment $k_{1}$ has the greatest influence on the PDF.  It significantly impacts both the peak cycle amplitude and the width of the distribution, making it the most critical factor in determining the overall behavior.  The result of $0.75\times k_{1}$ is more consistent with the reconstructed PDF of \citet{2014A&A...562L..10U} in terms of both peak amplitude and peak probability density.  The other parameters primarily affect the width, and have less influence on the peak amplitude.  A smaller $quench$ or larger $stoch$ results in a wider peak, while having a minor effect on the peak cycle amplitude.  We note that, since the stochastic component is multiplicative to $\Delta DM$, the actual scattering range is generally larger for larger $k_{1}$ value, and hence a wider PDF curve.

Knowing how the PDF is affected by the parameters, we can then show how the nonlinearity and stochasticity in reality may be different from that of J20.  We modify the parameters and look for a closer match to \citet{2014A&A...562L..10U}.  The close match shown by the orange curve in Figure \ref{fig:pdfs}d is obtained by modifying the parameters together.  This indicates that compared to results of J20, the B-L type dynamo may saturate at higher cycle amplitudes, but the saturated dipole moment should be smaller.  Meanwhile the stochasticity might be a little less.  For simplicity, this parameter set is made by some trials instead of parameter optimization, so it may not be the best fit, or the most realistic fit.   Still, it should be warned that the reconstructed sunspot numbers by \citet{2014A&A...562L..10U} has uncertainty as well, which may affect the evaluation of our results.

%The PDF we generate is a single component distribution.  Parameters affect its peak and shape.  (Fig \ref{fig:pdfs} a-c)

%Jiang2020's parameters can be modified a bit to match observations. (Fig \ref{fig:pdfs} d)

%ratio of "strong" cycles, an indicator of grand minima
%standard set of parameters: 0.25 +-0.01
%optimized set: 0.037 +-0.005
We would like to further note that the results above are about normal cycles.  Especially weak cycles (those weaker than the dashed line in Figure \ref{fig:pdfs}) are not described by our model.  For especially strong cycles, which are usually regarded as grand maxima, they appear as a tail of the distribution of the normal cycles instead of distinctive peak, which is similar to \citet{2014A&A...562L..10U}.  Hence, the question of whether these strong cycles are well described by the model is just the same problem of whether the normal cycles are well described.  In \citet{2014A&A...562L..10U}, strong cycles are approximated as the $+1\sigma$ range of the main normal cycle mode.  Converting to our definition of cycle amplitudes, this refers to cycles stronger than 258.  We use this to obtain quantitative results of the grand maxima phase.

We use 1000 cycles as a set and calculate the ratio of strong cycles over all cycles, and obtain the average and standard deviation from 1000 sets.  For the standard set of parameters, 0.25 of the cycles belong to the grand maxima phase, with the standard deviation being 0.01; while for the optimized set of parameters, the ratio is 0.037 with standard deviation of 0.005.  This result is a natural extension of the general property of the PDF of cycle amplitudes described above, and is also heavily affected by the parameters of nonlinearity and stochasticity.

\section{Implications for solar cycle prediction}\label{sec:disc}
%Remember to change cycle 25
%The iterative map of cycle amplitudes can be naturally used for cycle prediction, as long as the amplitude of the current cycle is known.

The iterative map of cycle amplitudes generates the amplitude of the next cycle, along with its associated uncertainty, when the current cycle amplitude is given.  This can be regarded as cycle prediction of the next cycle. The cycle amplitude can be obtained when the cycle has reached its maximum. If we use the maximum International Sunspot Number 2.0, and suppose that cycle 25 is already close to maximum, which is approximately 150, we can then produce a prediction of cycle 26.  Using Equation (\ref{eq:recursion2}) and the standard set of parameters, we get that the amplitude of cycle 26 will be 255 with $1\sigma$ uncertainty being 69.  If we use the parameter set in Figure \ref{fig:pdfs}d, which is more close to the observational PDF, the amplitude of cycle 26 will be 194 with $1\sigma$ uncertainty being 44.  This implies that iterative map can provide a efficient and simplified means for cycle prediction, only if the properties of the dynamo can be realistically quantified.

%\begin{figure}
%%\plotone{}
%gridline{\fig{observational.pdf}{0.5\textwidth}{}          }
%\caption{$SN(n)$ against $SN(n+1)$ for cycles 1 to 24 from %International Sunspot Number 2.0.  Each red diamond represents %an iteration from n to n+1.  The solid black curve is Equation %\ref{eq:recursion2}, with the dashed curves representing its %2$\sigma$ range.  Cycle amplitudes below 107.1 (cycle 5, 6) are %not included.\label{fig:obscycles}}
%\end{figure}

Our results show that the stochasticity is the primary source of cycle variability, and a realistic quantity of stochasticity is needed to form a PDF of cycle amplitudes similar to observations.  While the stochasticity may be weaker than the standard parameters we use, the difference should not be of several magnitudes.  Although the observational studies of the active region emergence is possibly still yet to be refined, it is certain that the stochasticity shall play an important role in cycle prediction.

Since stochasticity itself is intrinsically not predictable, it naturally limits the effective range of solar cycle prediction.  While predicting the next cycle at exactly the cycle minimum can be correct, prediction before the cycle minimum always has a uncertainty range which increases as we make prediction at earlier stage of cycle evolution, as shown by \citet{2018ApJ...863..159J}, because stochasticiy is involved.

\begin{figure}
%%\plotone{}
\gridline{\fig{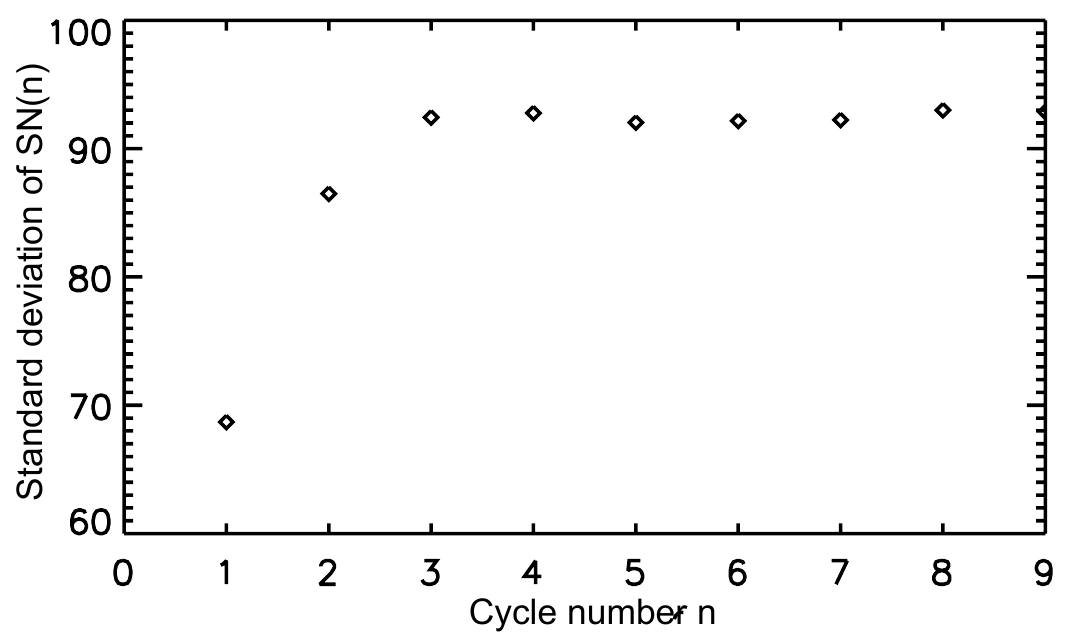}{0.5\textwidth}{}          }
\caption{The standard deviation of SN($n$) when SN(0) is 150. Each standard deviation is calculated from 10,000 SN($n$) values.\label{fig:predict}}
\end{figure}

For prediction beyond 1 cycle, the effect of stochasticity is even stronger, and long-term prediction is not valid from this perspective.  The recursion function itself already has a large uncertainty, and this inherent uncertainty affects cycle predictions at each iteration.  To give an example, we start at SN(0) = 150, which approximates the the amplitude of cycle 25, produce several sets of cycle amplitude series using the standard set of parameters, and calculate the standard deviations of SN($n$), which corresponds to the uncertainty of cycle prediction ahead of $n$ cycles.  As shown in Figure \ref{fig:predict}, the standard deviation of SN($n$) increases rapidly, and soon saturates after 3 or 4 cycles, where the standard deviation agrees with the cycle amplitude variations illustrated by the PDF in Figure \ref{fig:pdfs}.  This indicates that in our model, the uncertainty of long-term cycle prediction is too large for it to be meaningful.  A solar cycle quickly loses all of its original information after a few iterations and becomes completely indistinguishable from any solar cycle among the entire PDF of cycles. This loss of distinction is a direct consequence of stochastic processes.

\section{Discussions}\label{sec:discuss}
The iterative map is a well-established tool in the field of nonlinear dynamics. It facilitates the analysis of solar cycle variability and helps us explore the nature of the dynamic system governing solar dynamo processes. Despite its simplicity, the nonlinear behaviors it reveals are generic \citep{May1976}. Using the BL-type dynamo, \cite{2005ApJ...619..613C} confirmed that low-order dynamical systems can capture essential features of the full problem. The key challenge lies in realistically constructing the nonlinearity and stochasticity that characterize the solar cycle system.

\citet{Durney2000,Charbonneau2001} once presented iteration maps of solar cycle amplitudes. They demonstrated that, for certain parameter ranges, the solar dynamo could enter a chaotic regime, which contrasts with our findings. The discrepancy stems from the different forms of nonlinearity used. In the earlier studies, functions with free parameters were introduced to represent nonlinearity, without clearly specifying their physical origin. In contrast, we adopt forms of nonlinearity and stochasticity with well-defined interpretations and quantifications, based on the work of J20. This approach marks a clear advancement over previous studies. Other reason for the discrepancy lies in how the poloidal field, generated by the toroidal field of cycle $n$ (denoted as $\Delta DM$), is treated. In the earlier studies, $\Delta DM$ was mistakenly assumed to be equivalent to the dipole moment strength at the end of cycle $n$, i.e., $\Delta DM = DM\left(n\right)$. In reality, $\Delta DM$ must first cancel out the residual poloidal field from the previous cycle, meaning it should be expressed as $ \Delta DM = DM\left(n\right) + DM\left(n-1\right)$. The two major differences lead to fundamentally distinct interpretations of solar cycle variability when analyzed through iterative maps.

Our findings, which indicate the absence of chaos and the stochastic origin of solar cycle variability, agree with \citet{Cameron2017}, who attribute solar cycle variability to a weakly nonlinear limit cycle affected by random noise. Comparing to the model of \citet{Cameron2017} considering a B-L dynamo in general, we introduce the nonlinearity and stochasticity based on observational latitude and tilt quenching of active regions analyzed in J20. Such observation based nonlinearity cannot produce chaos regardless of parameters.  Generally, any $\alpha$-effect following an increase-then-saturate form cannot produce chaos. It is the similar decay properties of solar cycles in sunspot numbers that mainly leads to the saturate form of the $\alpha$-effect \citep{Cameron2016, 2018ApJ...863..159J, Biswas2022}.  While it has been shown that stochastic dynamo models can produce certain properties of cycle variability in previous works \citep[e.g.,][]{2000ApJ...543.1027C,2012PhRvL.109q1103C,2013ApJ...777...71O,2014A&A...563A..18P,2017ApJ...847...69K}, we clearly show that why nonlinearity should not be the origin of cycle variability in a generic method.

In our model, there is no modulation beyond 1 cycle, and it does not present long-term memory.  Successful cycle predictions for the weak cycle 24 using precursor methods \citep{2005GeoRL..32.1104S,2005GeoRL..3221106S} and solar dynamo models \citep{2007PhRvL..98m1103C, 2007MNRAS.381.1527J} have suggested that the memory of solar dynamo model is limited to within 1 cycle. The predictions of a stronger cycle 25 compared to cycle 24, made by surface flux transport based models \citep[e.g.,][]{Cameron2016b, 2018ApJ...863..159J, Jiang2018JASTP} and dynamo based models \citep[e.g.,][]{Bhowmik2018, Guo2021} also provide supporting evidence.  \citet{2021ApJ...913...65K} show that the memory of the polar field is determined by the dynamo regime.  Meanwhile, power spectra derived from the 1,000,000 solar cycle series generated based on the iterative map show no significant peaks around 8 cycles, which corresponds to the Gleissberg period \citep{Gleissberg1939}. This aligns with the finding of \citet{2019A&A...625A..28C}, who have also shown that long-term modulation is not needed to explain the solar cycle variability when stochastic forcing is present. Based on this, we argue that the iterative map, which lacks long-term memory, is a reasonable representation of solar cycle amplitude evolution, and longer-than-1-cycle modulations are not a primary concern as we evaluate the effects of nonlinearity and stochasticity.  As for the G-O rule, our model possesses G-O rule behavior when the cycle series are paired up.  More pairs tend to be constructed by a weaker cycle followed by a stronger cycle instead of the opposite.  This phenomenon is a result of the PDF that the cycles follow as shown in Figure \ref{fig:pdfs}, and the recursion function Equation (\ref{eq:recursion2}).  We will focus on the G-O rule and provide a more detailed examination and clearer definitions of it in the subsequent paper in this series \citep{Wang2025}.

The correlation coefficient between cycle $n$ and $n$+1 have been evaluated by observational studies to analyze the properties of cycle evolution.  Obviously, increasing nonlinearity and stochasticity both reduce the Pearson's correlation. Observationally, the correlation from directly observed cycles is rather weak and is not of statistical significance \citep[e.g.][]{2015SoPh..290.1851H}, although being somewhat positive.  The correlation in our model tends to be negative.  Such difference is probably a result of the limited number of cycles.  When the number is limited, the distribution of the correlation coefficient will become large, and positive values will be possible. The effect of limited number of solar cycles on the statistical properties of solar cycles will be addressed in the next paper of this series \citep{Wang2025}.

Nonlinearity and stochasticity are key factors controlling the probability density distribution of cycle amplitudes. The mode of the distribution is primarily determined by the maximum dipole moment generated by active regions, while the width of the distribution is influenced by both nonlinearity and stochasticity. Uncertainties in the parameters of nonlinearity and stochasticity arise from the limitations of observational constraints on the B-L type dynamo, posing challenges when comparing with observations. As summarized by \citet{2021A&A...653A..27J}, previous studies on tilt quenching have shown inconsistencies due to difference in datasets and methods.  Such inconsistencies occur on various perspectives of tilt angles, for example, the relationship between tilt angle and total flux or maximum flux density (Bmax).  \citet{2020ApJ...889L..19J,2024ApJ...966..112S} suggest that the tilt angle initially increases as Bmax increases but then decreases beyond a certain value.  \citet{2018ApJ...867...89L} suggests a slight anti-correlation between tilt angle and flux, while \citet{2012ApJ...745..129S} do not observe correlation between tilt angle and flux.  Recently, \cite{Qin2025} decrease measurement errors in tilt angles through mutual validation of datasets, enabling a more accurate analysis of their intrinsic properties. Although dynamo quenching due to AR properties is becoming clearer, further investigation is still needed.  Meanwhile, \citet{Petrovay2020JSWSC,Wang2021} suggest that near-equator meridional flows and surface turbulent diffusion also influence the contribution of ARs to solar cycle evolution. Variations in AR and flow properties will yield different parameters in our model, affecting the distribution of cycle amplitudes. In addition to uncertainties in the observed nonlinearity and stochasticity of B-L type dynamo, uncertainties in reconstructed sunspot numbers also limit the statistical conclusions.

While the recursion function for normal cycles is subject to parameter uncertainty, it is completely unconstrained for grand minima.  The observational constraints on B-L type nonlinearity and stochasticity during grand minima are extremely limited, for the ARs are scarce.  The generation of dipole moment for very weak cycles is lacking in the function we adopt, so a secondary peak for grand minima is missing in our probability density distribution, different from \citet{2014A&A...562L..10U}.  There are previous works suggesting that grand minima can be generated by stochasticity such as, \citet{2012PhRvL.109q1103C,2013ApJ...777...71O,2017ApJ...847...69K}, which are examples of producing long-term modulation of solar cycles from stochastic model.  However, their analyses of the grand minima mainly focus on the total time of grand minima compared to all cycles, and do not cover whether or not the PDF of cycles is double-peaked as presented by \citet{2014A&A...562L..10U}.  From this perspective, the origin of grand minima is not clear at present.  Better treatment of  weak cycles in the iterative map may produce better distribution containing more realistic grand minima in future work.

Grand maxima on the other hand, appear as tail of the distribution of normal cycles in our model, which is in agreement with observational results that do not find statistically significant grand maxima phase such as \citet{2014A&A...562L..10U,2018A&A...615A..93W}. In fact, at present grand maxima are not as well defined as grand minima, and it is not clear whether they have a distinct physical origin or not, as summarized by \citet{2023LRSP...20....2U}. The precise physical origin of grand maxima remains an open question for future studies.

\section{Conclusion}\label{sec:outro}
In this article, we produce an iterative map of solar cycle amplitude using the B-L type nonlinearity and stochasiticity based on observational properties of active regions.  We use the iterative map to generate and analyze solar cycle amplitude series.  We show that deterministic chaos is absent in our model regardless of model parameters, and stochasticity is always required to obtain solar cycle variability, which naturally limits the cycle prediction range.  The distribution of synthetic cycles resembles the observed normal cycles, leaving the grand minima to be included in the model.  The parameters of nonlinearity and stochasticity have profound influences on the distribution, and it can be inferred that the saturation of the B-L mechanism may have lower dipole moment saturating at higher cycle amplitude, as well as less stochasitcity compared to J20.  Our result implies that with the amplitude of an ongoing cycle, the following cycle can be predicted but with uncertainty range, and that long-term prediction ahead of several cycles quickly loses its validity completely.

%Compared to early iteration maps of cycle amplitudes such as \citet{Durney2000} and \citet{Charbonneau2001}, in our model the poloidal field generated by the toroidal field $\Delta DM$ needs to cancel out the poloial field of the previous cycle first, which leads to a critical result.  %This means that, as long as the generation of poloidal field increases monotonially and then saturates to a certain value, the fixed point is always stable if there is one.  Hence, the absence of chaos is not limited to the specific form of $\Delta DM$ we adopt from J20.

Generally, deterministic chaos cannot be present as long as, 1) the $\Omega$-effect is mostly linear; 2) the Hale's polarity law is preserved; and 3) the generation of poloidal field follows a increase-then-saturate form.  Furthermore, these requirements are not limited to Babcock-Leighton mechanism.  Other type of quenched $\alpha$-effects following these characteristics should also be devoid of chaos, and thus rely on stochasticity to generate varying cycle amplitudes.  The reason is that if there is a fixed point at the saturation of the $\alpha$-effects, the derivative of the recursion function always promotes its stability.  In contrast, if the B-L mechanism or any form of $\alpha$-effect actually decreases after reaching its peak with the increase of cycle amplitude as presented by \cite{2005ApJ...619..613C}, and the fixed points falls to the decreasing part, then chaos will be possible in the iteration map.  This result can provide important guideline to future nonlinear solar dynamo model studies.

%\clearpage

%Solar cycle variability requires stochasticity while nonlinearity along is not sufficient, as chaos cannot occur.  This is valid as long as the Delta DM is monotonically increasing, hence this result covers many type of dynamo models in which the quenching on alpha effect only saturates instead of actually decreases.

%The distribution of cycles is single component at present, and roughly resembles the normal cycle mode of observations.  Parameters have a profound effect on PDF, and some changes in parameters can produce a result resembling observations.

%Discussions: further studies on cycle evolution and dynamo can produce improved parameters.

%Other sun-like stars.

%The grand minima to be included.
\begin{acknowledgements}
    The authors acknowledge the utilization of the reconstructed sunspot number series by I. G. Usoskin, et al. (2014) (http://dx.doi.org/10.26093/cds/vizier.35629010), and the reconstructed sunspot number series by I. G. Usoskin, et al. (2021) (http://dx.doi.org/10.26093/cds/vizier.36490141).  The International Sunspot Number version 2.0 is from the World Data Center SILSO, Royal Observatory of Belgium, Brussels, which is available at https://www.sidc.be/SILSO/datafiles.  This research was supported by the National Natural Science Foundation of China through Grant Nos. 12425305, 12350004, 12173005, 12373111, \& 12273061, the National Key R\&D Program of China through Grant No. 2022YFF0503800, and supported by Specialized Research Fund for State Key Laboratory of Solar Activity and Space Weather. J.J. acknowledges the International Space Science Institute Team 474 for stimulating discussions.
\end{acknowledgements}

\bibliography{ref}{}
\bibliographystyle{aasjournal}
\end{document}